\documentclass{sig-alternate-05-2015}

\usepackage[T1]{fontenc}
\usepackage{lmodern}      
\usepackage[utf8]{inputenc}  
\usepackage{microtype}      
\usepackage{enumitem}
\usepackage{microtype}   
\usepackage{flushend}
\usepackage{ellipsis}   
\usepackage{mdframed}
\usepackage{graphicx}
\usepackage[draft]{hyperref}
\usepackage{subfig}
\usepackage{verbatim}
\usepackage{url}
\usepackage{float}
\usepackage{enumerate}
\usepackage{multicol}
\usepackage{multirow}
\usepackage{csquotes}
\usepackage{booktabs}
\usepackage{array}
\usepackage{fixltx2e}
\usepackage{listings}
\usepackage{balance}
\usepackage{mathtools}
\usepackage[colorinlistoftodos]{todonotes}

\usepackage{breqn}

\usepackage{graphics}

\let\svtodo\todo\renewcommand\todo[1]{\svtodo[inline]{#1}}

\makeatletter
\newcommand\footnoteref[1]{\protected@xdef\@thefnmark{\ref{#1}}\@footnotemark}
\makeatother

\newcolumntype{L}[1]{>{\raggedright\let\newline\\\arraybackslash\hspace{0pt}}m{#1}}
\newcolumntype{C}[1]{>{\centering\arraybackslash} m{#1} }
\newcolumntype{W}[1]{>{\centering\arraybackslash}m{#1}}

\begin{document}

% Copyright
%\setcopyright{acmcopyright}
%\setcopyright{acmlicensed}
%\setcopyright{rightsretained}
%\setcopyright{usgov}
%\setcopyright{usgovmixed}
%\setcopyright{cagov}
%\setcopyright{cagovmixed}

% DOI
%\doi{10.475/123_4}

% ISBN
%\isbn{123-4567-24-567/08/06}

%Conference
%\conferenceinfo{PLDI '13}{June 16--19, 2013, Seattle, WA, USA}

%\acmPrice{\$15.00}

%
% --- Author Metadata here ---
%\conferenceinfo{WOODSTOCK}{'97 El Paso, Texas USA}
%\CopyrightYear{2007} % Allows default copyright year (20XX) to be over-ridden - IF NEED BE.
%\crdata{0-12345-67-8/90/01}  % Allows default copyright data (0-89791-88-6/97/05) to be over-ridden - IF NEED BE.
% --- End of Author Metadata ---

%\title{Emotions in Jira -- Valence, Arousal, and Dominance}
\title{Mining Valence, Arousal, and Dominance -- Possibilities for Detecting Burnout and Productivity?}
%
% You need the command \numberofauthors to handle the 'placement
% and alignment' of the authors beneath the title.
%
% For aesthetic reasons, we recommend 'three authors at a time'
% i.e. three 'name/affiliation blocks' be placed beneath the title.
%
% NOTE: You are NOT restricted in how many 'rows' of
% "name/affiliations" may appear. We just ask that you restrict
% the number of 'columns' to three.
%
% Because of the available 'opening page real-estate'
% we ask you to refrain from putting more than six authors
% (two rows with three columns) beneath the article title.
% More than six makes the first-page appear very cluttered indeed.
%
% Use the \alignauthor commands to handle the names
% and affiliations for an 'aesthetic maximum' of six authors.
% Add names, affiliations, addresses for
% the seventh etc. author(s) as the argument for the
% \additionalauthors command.
% These 'additional authors' will be output/set for you
% without further effort on your part as the last section in
% the body of your article BEFORE References or any Appendices.

\numberofauthors{1} 
\author{\alignauthor Mika M{\"a}ntyl{\"a}$^{\text{1}}$, Bram Adams$^{\text{2}}$, Giuseppe Destefanis$^{\text{3}}$, Daniel Graziotin$^{\text{4}}$, Marco Ortu$^{\text{5}}$\\
\affaddr{$^{\text{1}}$M3S, ITEE, University of Oulu, Finland}, \url{mika.mantyla@oulu.fi}\\
\affaddr{$^{\text{2}}$MCIS, Polytechnique Montreal, Canada}, \url{bram.adams@polymtl.ca}\\
\affaddr{$^{\text{3}}$Brunel University London, Uxbridge, United Kingdom}, \url{giuseppe.destefanis@brunel.ac.uk}\\
\affaddr{$^{\text{4}}$Free University of Bozen-Bolzano, Italy; University of Stuttgart, Germany}, \url{daniel.graziotin@unibz.it}\\
\affaddr{$^{\text{5}}$DIEE, University of Cagliari, Italy}, \url{marco.ortu@diee.unica.it}
}

\CopyrightYear{2016}
\setcopyright{acmlicensed}
\conferenceinfo{MSR'16,}{May 14-15 2016, Austin, TX, USA}
\isbn{978-1-4503-4186-8/16/05}\acmPrice{\$15.00}
\doi{}

\maketitle

\begin{abstract}
%\todo{rewrite (reduce to half of size) in terms of burn-out, VAD more direct information than discrete, exploratory study using VAD on issue reports}
Similar to other industries, the software engineering domain is plagued by psychological diseases such as burnout, which lead developers to lose interest, exhibit lower activity and/or feel powerless. Prevention is essential for such diseases, which in turn requires early identification of symptoms.
 %For this reason, ongoing research on human emotion detection has established 
The emotional dimensions of Valence, Arousal and Dominance (VAD) are able to derive a person's interest (attraction), level of activation and perceived level of control for a particular situation from textual communication, such as emails. 
%Since automatic monitoring of VAD metrics from software development repositories could be a promising way to identify project members that require psychological support, %The dimensional approach of emotions claims that all experiences of human emotions can be mapped to the three-dimensional space of Valence and Arousal and Dominance (VAD). Valence is the attractiveness, Arousal represents the activation, and Dominance represents who is perceived to control the situation. We hope that in the future VAD could help to identify developers with high risk of burnout and explain differences in productivity.  
As an initial step towards identifying symptoms of productivity loss in software engineering, this paper explores the VAD metrics and their properties on 700,000 Jira issue reports containing % software issue tracking systems using two open datasets. The first dataset consists of roughly 700,000 Jira issues with
over 2,000,000 comments, since issue reports keep track of a developer's progress on addressing bugs or new features. %We measure VAD in these comments using % . The second dataset comes from psychology and provides 
Using a general-purpose lexicon of 14,000 English words with known VAD scores, 
our results show that issue reports of different type (e.g., Feature Request vs. Bug) have a fair variation of Valence% can be mapped to issue Type
, while increase in issue priority (e.g., from Minor to Critical) typically increases Arousal. % differs between issue reports of different priority% can be mapped to issue Priority
%We find evidence that these emotions are contagious within an issue life cycle. 
Furthermore, we show that as an issue's resolution time increases, so does %is in the process of being resolved, %
 the arousal of the individual the issue is assigned to. Finally, the resolution of an issue increases valence, especially for the issue Reporter and for quickly addressed issues. The existence of such relations between VAD and issue report activities shows promise that text mining in the future could offer an alternative way for work health assessment surveys. 
\end{abstract}

\section{Introduction}

%Affects (i.e., emotions and moods such as joy, anger or sadness)\footnote{\label{see-more}In this paper, we use the terms affect and emotions interchangeably, in line with many authors~\cite{Graziotin2015}.
%}
Emotions\footnote{\label{see-more}In this paper, we use the terms affect and emotions interchangeably, in line with many authors~\cite{Graziotin2015}.} and moods, such as joy, anger or sadness,
pervade the daily operations of organizations. Emotions are the primary drivers of employees when dealing with deadlines, motivation to work, sense-making, human-resource processes, behaviour, and work performance~\cite{Barsade2007}. For example, Graziotin et al.~\cite{graziotin2013happy} and Khan et al.~\cite{khan2011moods} adopted the Valence-Arousal-Dominance (``VAD'') affect representation to conceptualize an individual's emotional spectrum% using three dimensions--or variables
. These studies showed a positive relationship between software developers' productivity and how they enjoyed a situation (high ``Valence'') and were feeling in control of the development task (high ``Dominance''). In psychology and management, it is generally accepted that increased alertness or readiness to act (high ``Arousal''), improves employees' performance (typically because of time pressure or reward-punishment schemes). This has been shown to apply to software engineering as well~\cite{nan2009impact,mantyla2014time}.

Yet, changes in affect in terms of VAD can also have adverse effects. Increases in Arousal start to hamper performance from a certain threshold (Yerkes-Dodson law~\cite{YerkesDodsonLaw}), with Arousal caused by increased and prolonged pressure even leading to burnout in software teams~\cite{sonnentag1994stressor}. %For example, from a certain threshold, increases in Arousal start to hamper performance, also known as the Yerkes-Dodson law~\cite{robert1908relation}, with Arousal caused by increased and prolonged pressure even leading to burnout in software teams~\cite{sonnentag1994stressor}. 
This is why % To address pressure-related problems at work,
major IT companies like Google have promoted various remediation techniques such as mindfulness training~\cite{tan2012search}. It has also been shown that a high need for independence, which can be linked to Dominance, is one of the factors characterizing software engineers~\cite{beecham2008motivation}. Hence, lack of such independence or control at work, increases the risks of burnout in software development~\cite{sonnentag1994stressor}. In other words, studies on the emotions of VAD dimensions in software engineering are important as they possibly could identify symptoms of high productivity, i.e., when someone experiences high valence, dominance and arousal, but also symptoms of where the risk of burnout increases, i.e., when a person experiences low valence, low dominance and high arousal. 

Generally speaking, emotions have been studied in psychology either using a discrete approach or a dimensional approach~\cite{Graziotin2015}. The discrete approach represents emotions as a set of basic affective states that can be distinguished uniquely such as anger, joy, sadness, and love~\cite{Plutchik1980}. The dimensional approach (proposed as early as 1897), groups affective states in a smaller set of major dimensions, e.g., VAD~\cite{wundt2004outlines}.  Thus far, past studies on mining affects from software repositories, e.g.,~\cite{Murgia2014}, have focused almost exclusively on utilizing discrete emotion theories. We think employing a dimensional approach (VAD) is more advantageous than using the discrete approach as the dimensional can be linked to burnout and productivity as discussed in the previous paragraphs.

The few studies adopting a VAD approach in software engineering research, e.g.,~\cite{graziotin2013happy, khan2011moods, Muller2015}, primarily have observed, measured or queried humans in experimental or quasi-experimental settings while working on software engineering tasks. Instead, this paper mines software repositories using the VAD approach, as this allows automatic, non-intrusive, retrospective, real-world assessment of VAD % utilizing a big data approach over
across several thousands individuals (instead of small sample sizes) to determine existing relations between VAD and issue report productivity. We study VAD in issue reports, since in many open source projects these represent the daily communication medium to discuss ongoing work, failures and successes, and these reports are also where end users and developers meet each other. Hence, the issue repository is a representative software repository for the study of VAD.%Additionally, due to the large volume of analyzed data we are less likely to be affected by random variation than past works. 
%\todo{Daniel: Similarly here, not first step in general. I added ``mining'' here. Please check}
%As a first step towards exploring the role of VAD mining in software repositories, this paper uses % In this paper, we demonstrate the use VAD dimensions in software engineering by utilizing 

Using the largest available lexicon of VAD% Valence (attractiveness of an event, object, or situation), Arousal (activation level towards a stimulus, mental awakening, and alertness to responses), and Dominance (perceived domination of a stimulus to the subject)
, containing 13,915 English words% , provided by Warriner et al.
~\cite{Warriner2013}, we analyze the presence of VAD in the % Apache \todo{this is not right, there are other sources as well?} 
Jira issue report data set of % one of the largest datasets of issues reports openly available by
Ortu et al.~\cite{Ortu2015}, which contains 700,000 issues of 1,000 open source projects, including two million issue comments. 
Given the lack of ``gold standard'' for describing developers’ real emotions, we explore the data by making reasonable assumptions on how emotions should change, e.g., getting an issue resolved should increase Valence. In particular, we address the following research questions:

\textbf{RQ1: How does VAD relate to issue report characteristics?} Theoretically, we expect issues with higher priority to be linked to higher Arousal (developers are more active). Furthermore, Valence should be low for defects (less attractive), while higher for new features and improvements. Finally, issues that are fixed swiftly likely featured developers with high Dominance (i.e., who felt in control).%, more pleasure is related to items that are new than old  and less pleasure is related to things that are not working, i.e. bugs. 

\textbf{RQ2: How does VAD evolve when issues are resolved?} Since issue reports are conversations between the initial reporter of a bug or new feature and developers, and can last for a long time, the initial VAD values could be different at the start compared to the end of the discussions. For example, one would expect low Valence initially for a bug, but higher Valence when the bug is subsequently fixed.% should be seen for fixed defects when they are resolved, e.g. last comments

\textbf{RQ3: Can VAD explain the time used for fixing a defect?} We compare the impact of VAD measures on defect fixing time to that of discrete emotion, sentiment and politeness measures~\cite{Ortu2015a}. Given the higher-level information provided by VAD measures, we expect to find a stronger impact for them. However, as we use a general-purpose lexicon, we might also experience lower impact than Ortu et al.~\cite{Ortu2015a}, who used measures collected by software engineers.

\textbf{RQ4: What issue characteristics predict VAD scores in the issue comments?} Since the cause-effect relationship between issue properties and VAD can also be argued to go the other way, we check what issue properties affect individuals' emotions toward an issue. For example, which issues are the ones giving the most pleasure (Valence) and which ones give the most stress (Arousal)?

\section{Background and Related work}

Affect has been defined as ``any type of emotional state {[}\ldots{}{]} often used in situations where emotions dominate the person's awareness''~\cite{VandenBos2013}, and it has been used as an umbrella term for emotions and moods~\cite{Plutchik1980}\footnoteref{see-more}. Emotions have been studied and represented through several theoretical models, which have been classified using either a discrete approach or a dimensional approach~\cite{Graziotin2015}. While the discrete approach represents emotions as a set of basic affective states that can be distinguished uniquely~\cite{Plutchik1980} (``that person is angry''), the dimensional approach groups affective states in a smaller set of major dimensions (``that person has a Valence of X and Arousal of Y''), see Graziotin et al.~\cite{Graziotin2015}. The discrete approach is useful when attempting to study % particular emotions as intervals of the affective spectrum of individuals. That is, when there is the aim of studying some 
particular emotions--say frustration, anger, and fear% --it is sensible to employ the discrete approach
. When the aim is to study all emotions and moods expressed by individuals, the discrete approach is limited to only those emotions defined by the chosen theory% . The discrete framework has does not cover the entirety affective spectrum of individuals
, making it difficult to link to interesting outcome variables such as the risk of burnout or high productivity \cite{Lutz1986}. Additionally, it has been shown that, in contrast to the discrete emotion framework, the VAD measures are independent from any cultural or linguistic interpretation~\cite{guerini2015deep,Russell1991}, which makes VAD more robust.

\emph{Valence} is the emotional dimension related to the attractiveness (or adverseness) of an event, object, or situation~\cite{Lewin1935,Lang1993}. The term refers to the ``direction of a behavioral activation associated toward (appetitive motivation) or away (aversive motivation) from a stimulus''~\cite{Lane1999}. \emph{Arousal} is the dimension representing the emotional activation level~\cite{Lane1999}. It has various physiological and psychological responses, e.g., an increased heart rate and alertness to responses, and it is perceived as a sensation of being reactive to stimuli and mentally awake, i.e., vigor and energy or fatigue and tiredness~\cite{Zajenkowski2012}. Arousal also intensifies the pleasure or displeasure described by the Valence dimension~\cite{Storbeck2008}, e.g., frustration can change to anger and content can change to delight when Arousal increases~\cite{Graziotin2015}. Finally, \emph{Dominance} represents a change in the sensation of having control on a stimulus (or a situation)~\cite{Lang1994}.% , see Figure 1 in

The foundation of affect mining resides in psychology studies that map individual words to affects. The first and most cited study was performed by Russell~\cite{Russell1980}, who used 36 subjects to map 28 stimulus words to a Valence-Arousal space named the circumplex model of affect. Since the work of Russell, the amount of words that have been mapped to the Valence-Arousal space has increased considerably. Recently, Warriner et al.~\cite{Warriner2013} used 1,865 participants to rate 13,915 English words to create the largest collection of individual words mapped to a Valence-Arousal-Dominance space.

Even though, in the past ten years, openly available software repositories have been vital for boosting empirical research in software engineering, the field of affect mining in software repositories is still emerging, and has not considered VAD thus far. %and some articles even claim that the related research in the area is pretty much non-existing ~\cite{Short2015,Jurado2015} .
Guzman et al.~\cite{Guzman2013b,Guzman2013c} have proposed prototypes and initial descriptive studies towards the visualization of affect over a software development process. In their work, the authors applied sentiment analysis to data coming from mailing lists, web pages, and other text-based documents of software projects. Guzman et al. built a prototype to display a visualization of the affect of a development team, and they interviewed project members to validate the usefulness of their approach.
In another study, Guzman et al.~\cite{Guzman2014}, performed sentiment analysis of Github's commit comments to investigate how emotions are related to a project's programming language, the commits' day of the week and time, and the approval of the projects. The analysis was performed over 29  top-starred Github repositories implemented in 14 different programming languages. The results showed Java to be the programming language most associated with negative affect. No correlation was found between the number of Github stars and the affect of the commit messages.

Begel et al.~\cite{begel2014using} investigated an approach to classify the difficulty of coding activities using psycho-physiological sensors (eye-tracker, electodermal activiy sensor and electroencephalography sensor), conducting an experiment with 15 professional developers.
Experimental results showed that it was possible (precision of over 70\% and a recall over 62\%) to train a Naive Bayes classifier on short or long time windows with a variety of sensor data to predict whether a new participant will perceive his tasks to be difficult. % The study demonstrated that is possible to use sensors to classify task difficulty. 

De Choudhury and Counts~\cite{DeChoudhury2013a} studied affect in 200k microblogging posts from 22k unique users of a Fortune 500 software corporation. 
The authors found that positive affects drop significantly in the evening with respect to the morning (but negative affects do also drop, although less significantly). Additionally, users that are central in the enterprise's network tend to share and receive highly positive affect, while those in individual contributor roles tend to express more negative affect.

Tourani et al.~\cite{Tourani2014} evaluated the usage of automatic sentiment analysis to identify distress or happiness in a development team.  The authors mined sentiment values from the mailing lists of two mature projects of the Apache software foundation considering both users and developers. The results showed that sentiment analysis tools obtained low precision on emails written by developers due to ambiguities in technical terms and difficulties in distinguishing positive or negative sentences from neutral. 

Murgia et al.~\cite{Murgia2014} studied whether issue reports carried emotional information about software development. The authors found, by manually analyzing the Apache Software Foundation issue tracking system, that developers did express emotions like sadness, joy and gratitude.

Jurado and Rodriguez~\cite{Jurado2015} gathered the issues of nine high profile software projects hosted on GitHub. Through an analysis of the occurrence of Ekman's~\cite{Ekman1992} basic emotions among the projects and issues, the authors discovered that in open source projects, sentiments expressed in the form of joy are % pervading and are present at 
almost one magnitude of order more common than the other basic emotions. Still, more than 80\% of the content was not classified as exhibiting a high amount of sentiment. Several other studies have been conducted using sentiment analysis and emotion mining for analyzing app reviews, in order to gain insights such as ideas for improvements, user requirements and to analyze customer satisfaction (e.g.,~\cite{guzman2014users,maalej2015bug}). 

Ortu et al.~\cite{Ortu2015a} analyzed the relation between sentiment, emotions and politeness of developers for Jira comments with the issue resolution time. The results showed that positive emotions and politeness were related to shorter issue fixing time. On the other hand, negative emotions were linked with longer issue fixing time.

%\subsection{Software engineering and repositories}

%%% Local Variables:
%%% mode: latex
%%% TeX-master: "paper"
%%% End:

\section{Method}

%To address the exploratory research questions of the introduction, we required a representative software repository data set on which to measure VAD measures as well as a representative lexicon with which VAD measures could be computed.

\subsection{Data Set}
\label{sec:data-set}

Given the potential of VAD to measure the impact of work-related events on productivity and mental health from textual communication, we opted to evaluate VAD on issue reports. Such reports correspond to defects and feature requests made by end users or developers. 
Typically, the reporter provides any relevant information, including a title and description of the defect or feature, after which interested developers or project members can comment, collaborate and review patches to fix the defect or implement the feature. 
While mailing lists contain communication related to usage issues, bugs and general topics, issue reports contain the day-to-day work assignments and communication. Hence, problems of motivation or signs of (hyper-)activity should be reflected in the natural language title, description and comments of an issue report.

As concrete data set, we selected Ortu et al.'s set of 700,000 issue reports of the Apache Foundation open source projects~\cite{Ortu2015}, spanning two million comments across one thousand projects. These projects use the popular Jira issue repository technology. For each report, we extracted title, description and comments, then calculated the VAD measures. % as outlined in the next section. 
We then used \textit{R}'s \textit{tm} package tokenizer to extract words, after which we matched the words found to the ones existing in our VAD lexicon (see below). If the word did not match our lexicon, then it was not used as no VAD score could be given to it. This mitigates the problem that issue reports can contain code and other information like stack traces.
 %, we used regular expressions to filter out such non-natural language before calculating any score~\cite{bacchelli2009benchmarking}.

\subsection{Measuring VAD}
\label{sec:measuring-vad}

All measures of VAD are based on a list of words that have manually been analyzed and assigned a VAD score. Warriner et al.'s~\cite{Warriner2013} leading lexicon contains 13,915 English words with VAD scores for Valence, Arousal and Dominance. To calculate the corresponding VAD scores for a piece of text (i.e., a list of words $\bar{w}=[w_1,w_2,...,w_n]$), the Range of the words' individual VAD scores is computed by taking the two words with the Max and Min Valence, Arousal or Dominance. For the special cases when Max has lower than average value or when Min has higher than average value we set the Max or Min to the average of all words of the lexicon ($\bar{W}=[W_1,W_2,...,W_N]$, where N is 13,915). Our formula is adapted from the one used by SentiStrength~\cite{thelwall10}, which is an industry strength tool to measure positive and negative sentiment. Our $avg(\bar{W})$ is similar to zero in SentiStrength, i.e. a mid-point and in SentiStrength Min values are never allowed to be higher than zero and Max values are never lower than zeros. In more formal terms, our equation is as follows:
\begin{dmath*}
Range(\bar{w})=\left\{
                \begin{array}{ll}
                  max(\bar{w})-avg(\bar{W}),\ if\ min(\bar{w})>avg(\bar{W})\\
                  avg(\bar{W})-min(\bar{w}),\ if\ max(\bar{w})<avg(\bar{W})\\
                  max(\bar{w})-min(\bar{w}),\ if\ min(\bar{w}) \leqslant avg(\bar{W}) \leqslant max(\bar{w})
                \end{array}
              \right.
\end{dmath*}

For example, if an issue would contain all the words listed in \autoref{table-mapping-words-va}, it would receive a Valence score of 5.81 (8.21-2.40, cf. third case in formula). The higher the value, the more extreme the VAD scores are. %A high value\bram{what does high or low value mean? i.e., low value could mean that all words have low Valence score or high Valence score, are both situations the same? i.e., what is difference between Valence score and Valence Range?}%\todo{Too dangerous?} 
Overall, our approach is simple and straightforward from a text-mining viewpoint.% and we provide avenues for it is improvement in Section 
%To calculate the Arousal of an issue report comment, one should check the Arousal score of each word in the lexicon, then average the values into one global Arousal score for the whole comment. 
% \todo{does this imply that that table has the score of the literal string ``Anger'' and not necessarily of ``any word related to the emotion anger''? Mika: No when people see those strings they obviously interpert them as words and emotions. Of course our approach has the weakness that we think that words people use actually represent their emotions}.

\begin{table}[t]
\renewcommand{\arraystretch}{1.3}
\caption{Mapping of the words representing discrete emotions to the Valence-Arousal-Dominance space% using Lexicon from
~\cite{Warriner2013}.}
\label{table-mapping-words-va}
\centering
\begin{tabular}{|l|l|l|l|}
\hline
        & Valence & Arousal & Dominance     \\ \hline
Anger   & 2.50    & 5.93 & 5.14 \\ \hline
Joy     & 8.21    & 5.55 & 7.00 \\ \hline
Sadness & 2.40    & 2.81 & 3.84 \\ \hline
Love    & 8.00    & 5.36 & 5.92 \\ \hline
\end{tabular}
\end{table}

% These scores are then further used to produce the results of this study.

%%% Local Variables:
%%% mode: latex
%%% TeX-master: "paper"
%%% End:

\section{Results}

This section discusses the motivation, approach and findings for each research question. 
The individual findings are then discussed in more depth in \autoref{sec:disc}.

\subsection*{RQ1: How does VAD relate to issue report characteristics?}

\subsubsection*{RQ 1.1. Is there a link between issue priority and Arousal?}

\begin{table}[t]
\renewcommand{\arraystretch}{1.3}
\caption{Interpretation of Cohen's d effect size.}
\label{tab:cohen}
\centering
\begin{tabular}{l|l}
$d<0.2$   & trivial effect size \\
$0.2\leq d<0.5$     & small effect size \\
$0.5\leq d<0.8$ & medium effect size \\
$0.8\leq d$    & large effect size \\
\end{tabular}
\end{table}

\textbf{Motivation.}  Arousal typically is introduced in a controlled experiment in the form of penalties~\cite{YerkesDodsonLaw}  and rewards~\cite{Ariely2009Stakes} that are awarded if a task can be completed within a certain time frame. In the context of issue reports, we believe that the reward for fixing a defect successfully is related to the priority of the issue, with high priority issues (e.g., blocker defects that are showstoppers) boosting a developer's profile and morale the most. % in the  issue priority represent the size of thereward in the software developers' world, i.e. the greater the priority of the issue the greater the reward in terms of improved product quality.
Hence, based on the psychological principles that cause Arousal, we expect to see increasing Arousal with increasing issue priority. 

\textbf{Approach.} For each issue element (i.e., for the title, description, first/last comment and all comments), we compared the Arousal score across the 5 issue priorities supported by Apache Jira. To do this, we grouped issue elements by their issue's priority% of  separately  grouped by issue priority (Apache Jira uses 5 different priorities)
, then performed t-tests to assess the statistical significance of each difference\footnote{\label{Bonf} Please note that when assessing significance, we had to adjust the alpha level using a Bonferroni correction. For example, as \autoref{tab:Arousalpriority} performed 20 comparisons, the normal significance level $\alpha=0.05$ turns into 0.0025 (0.05/20).} and compute the Cohen's \textit{d} for the effect size. The Cohen's \textit{d} effect size is a measure of how large a statistically significant difference is. \autoref{tab:cohen} shows how to interpret these \textit{d} values.

\textbf{Findings.} {\bf Blocker issues have higher Arousal than Trivial issues, but effect sizes are small.} \autoref{tab:Arousalpriority} shows how the Arousal score increases from Trivial to Blocker for all of the issue elements. However, there is no consistent increase from Critical to Blocker issues, indicating that the developers' Arousal level is not affected by this final difference in the priority scale. Furthermore, the effect sizes are negligible in between priorities, which means that even though differences are statistically significant, in practice the observed difference is not remarkable.% A more elaborate discussion of these and other findings is provided in Section~\ref{sec:disc}. 

If we compare the extreme priorities, we observe larger effect sizes. For example, the Arousal score of the issue description between Blocker issues (mean= 3.9541) and Trivial issues (mean=3.7299) has a Cohen's \textit{d} of 0.324, while the Arousal score of All comments between Blocker issues (mean=3.9771) and Trivial issues (mean= 3.7299) has a Cohen's \textit{d} of 0.354.

\begin{table}[t]
\centering
\caption{Arousal vs. issue priority. For each issue element, we show the mean (m) Arousal score, p-value and Cohen's d. The p and d values are always in comparison with the issue priority on the right.% effect size (Note: In each column the p and d are in comparison the next column in left )
}
\resizebox{\columnwidth}{!}{%

\begin{tabular}{ll|lllll}
&&Blocker       & Critical & Major   & Minor             & Trivial                                      \\
\hline
              & m     & 3.8512  & 3.8406            & 3.7814            & 3.7268            & 3.6668 \\
Title         & p        & 0.1364  & \textless 2.2e-16 & \textless 2.2e-16 & \textless 2.2e-16 &        \\
              & d        & 0.0134  & 0.0751            & 0.0696            & 0.0784            &        \\
\hline
              & m     & 3.9541  & 3.9609            & 3.8776            & 3.8434            & 3.7299 \\
Desc   & p        & 0.2626  & \textless 2.2e-16 & \textless 2.2e-16 &  \textless 2.2e-16          &        \\
              & d        & -0.0099 & 0.1173            & 0.04819            & 0.1609            &        \\
\hline
              & m     & 3.9771  & 3.9744            & 3.8903            & 3.8677            & 3.7428 \\
All  & p        & 0.6688  & \textless 2.2e-16 & \textless 2.2e-16 & \textless 2.2e-16 &        \\
              & d        & 0.0041 & 0.1233            & 0.0332            & 0.1843            &        \\
\hline
              & m     & 3.8161  & 3.8290            & 3.7688            &3.7281            & 3.6192 \\
First & p        & 0.0756  & \textless 2.2e-16 & \textless 2.2e-16 & \textless 2.2e-16 &        \\
              & d        & -0.0176 & 0.0833            & 0.0565            & 0.1528            &        \\
\hline
  & m     & 3.8047  & 3.8060            & 3.7690            & 3.7532           & 3.6767 \\
      Last        & p        & 0.8576  &3.608e-13          &3.145e-09         & \textless 2.2e-16 &        \\
              & d        & -0.0018  & 0.0514            & 0.0223            & 0.1086            &       
\end{tabular}%
}
\label{tab:Arousalpriority}
\end{table}

\subsubsection*{RQ 1.2. Is there a link between issue type and Valence?}

\textbf{Motivation.}  Valence refers to the pleasure or attraction experienced by humans. Typically, humans experience more pleasure in new things, which explains why hedonic shopping is a major driver in retail sales~\cite{arnold2003hedonic}. %For example, in the lexicon \cite{Warriner2013} we utilize the word "new" has higher Valence than the word "old" or "defect" , 7.68 vs. 3.19 and 2.89. 
Similarly, we hypothesize that in software engineering new features are associated with higher Valence than defect fixes, while defects are expected to be associated with negative emotions.% are expected to be associated to situations when something is defective.% broken and positive emotions are at play when something is new or being improved.   %\todo{Mika: Chicken egg problem due to SE terminology. 
%Needs to discussed howeever in previous Arousal scores for Blocker (=Block), Critical, Major, Minor, Trivial 4.25, 4.95, 4.24,  3.26,  3.82}

\textbf{Approach.} 	Since the data set contains multiple issue types, we focused on the top nine types, based on popularity. Given the expected similarities between different types in terms of Valence, we re-classified them into three groups, i.e., Bug, All Tasks, and Future Dev. The Bug group consists of Bug issue types only. The All Tasks group consists of Task, Sub-task, and Test issue types, which refer to regular development tasks. Finally, Future Dev consists of the Wish, New Feature, Improvement, Feature Request, and Enhancement issue types.

Hence, the Bug group represents all issues that refer to something being broken and we expect low Valence for it. The All Tasks group contains regular tasks for which neither low nor high Valence is expected. Finally, for the Future Dev group we expect high Valence as this is about creating new things rather than fixing old code. We then used the same analysis as for RQ 1.1.

\textbf{Findings.} {\bf Valence is lowest for Bugs, which supports our expectations.} As expected, \autoref{tab:Valencetype} shows that for all issue elements Valence is the lowest for Bugs. This supports the idea that developers would experience more pleasure from developing new features and other tasks in comparison to fixing bugs. Limited support was found for the conjecture that Future Dev tasks  bring more pleasure than other types of tasks (All Tasks). For issue title and issue description, we find small effect sizes, higher than 0.3, between All Tasks and Bug issue types.

\begin{table}[t]
\centering
\caption{Valence vs. issue type. For each issue element, we show the mean (m) Arousal score, p-value and Cohen's d. The p and d values are always in comparison with the issue type on the right.% effect size (Note: In each column the p and d are in comparison the next column in left )
}
\begin{tabular}{ll|lll}
&&Future Dev    & All Task & Bug                                          \\
\hline
    & m     & 5.9590            & 5.8724            & 5.4942 \\
     Title         & p        & \textless 2.2e-16 & \textless 2.2e-16 &        \\
              & d        & 0.0844            &0.3886            &        \\
\hline
   & m     & 5.8958            &5.9049            & 5.6527 \\
Desc              & p        & 0.0423 & \textless 2.2e-16 &        \\
              & d        & -0.0090            & 0.3447            &        \\
\hline
  & m     & 5.8346            & 5.8302            & 5.6362 \\
All              & p        & 0.3672 & \textless 2.2e-16 &        \\
              & d        & 0.0043            & 0.1927            &        \\
\hline
 & m     &5.9536            & 5.9120            &5.6714 \\
First              & p        & 7.165e-16 & \textless 2.2e-16 &        \\
              & d        & 0.0391            & 0.2212            &        \\
\hline
  & m     & 6.1023            & 6.0414            & 5.9042 \\
Last              & p        & \textless 2.2e-16 & \textless 2.2e-16 &        \\
              & d        & 0.0534            & 0.1176            &       
\end{tabular}
\label{tab:Valencetype}
\end{table}

\subsubsection*{RQ 1.3. Is there a link between issue resolution time and Dominance?}

\textbf{Motivation.}  Dominance refers to feeling in control of a situation. If a developer writing a comment to an issue report has a firm grip on the situation, this should be reflected in higher Dominance scores. Hence, we hypothesize that high Dominance in an issue report or comment should be associated with shorter resolution time, since if we are in control of the situation, we know what we are doing and likely will do it swiftly.

\textbf{Approach.}  We use an analysis similar to RQ 1.1. and RQ 1.2. Since issue resolution time is a continuous variable, we discretized it into ``high'' and ``low'' resolution time, each of which contains the elements of half of the issue reports.%half of the issue elements. For issue resolution time we splitted the data set from the middle to form two groups high and low issue resolution time. 

\textbf{Findings.}  {\bf To our surprise, high Dominance was associated with high (not low) issue resolution time.} Indeed, \autoref{tab:Dominanceresolution} shows a statistically significant difference but in the opposite direction than we were expecting, with higher Dominance associated with high issue resolution time in four out of the five cases studied. Only for the All Comments and Last Comment issue elements, we found that higher Dominance coincided with faster issue resolution. However, in all five cases the effect size is trivial and very close to zero. Hence, in practice the differences in resolution time do not seem to be remarkable.% Hence, only the last case matches our hypothesis.% and we are unable to explain the opposite direction in the results. 

\begin{table}[t]
\centering
\caption{Dominance vs. resolution time. For each issue element, we show the mean (m) Arousal score, p-value and Cohen's d. The p and d values are always in comparison with the resolution time on the right.% effect size (Note: In each column the p and d are in comparison the next column in left )
}
\begin{tabular}{ll|ll}
&&Short time    & High time                            \\
\hline
         & m      & 5.7582            & 5.7821 \\
Title              & p         & \textless 2.2e-16          &        \\
              & d         &-0.0278           &        \\
\hline
   & m      & 5.7544            & 5.8093 \\
Desc              & p         & \textless 2.2e-16 &        \\
              & d         & -0.0749           &        \\
\hline
  & m      & 5.7261            & 5.7268 \\
All              & p         &  0.7794 &        \\
              & d         &-0.0010           &        \\
\hline
 & m      & 5.7352            & 5.7888 \\
First              & p         & \textless 2.2e-16 &        \\
              & d         & -0.0678           &        \\
\hline
  & m      & 5.8979            & 5.8766 \\
Last              & p         & 8.766e-12 &        \\
              & d         & 0.0270            &       
\end{tabular}
\label{tab:Dominanceresolution}
\end{table}

\begin{table}[t]
\centering
\caption{Valence (V),  Arousal (A), and Dominance (D) between the first and last comment of all, assignees', reporters'  and others' comments of the analyzed closed issues. p is the significance of paired t-test and d is effect size Cohen's d}
\label{tab:1stLast_All}
%\begin{tabular}{m{1.3cm}m{0.2cm}|m{1cm}m{1.1cm}m{1.1cm}m{1.1cm}}
\begin{tabular}{ll|llll}
                           &   & All     & Assignees'        & Reporters'         & Others'           \\ \hline
\multirow{2}{*}{V}   & p & \textless 2.2e-16  & \textless 2.2e-16 & \textless 2.2e-16 & \textless 2.2e-16 \\
                                         & d & 0.37                           &  0.11 & 0.37           & 0.20            \\ \hline
\multirow{2}{*}{A}   & p & 7.67e-13  & \textless 2.2e-16 & 6.27e-05 & \textless 2.2e-16 \\
                                        & d & -0.0257                     & -0.151             & 0.036           & -0.097            \\ \hline
\multirow{2}{*}{D} & p & \textless 2.2e-16  &8.338e-09 & \textless 2.2e-16 & \textless 2.2e-16 \\
                                         & d & 0.28                            & 0.045            & 0.28            & 0.21           
\end{tabular}
\label{tab:1stLast}
\end{table}

\subsubsection*{Summary}
\begin{figure*}[t]
\centering
\subfloat[]{\label{fig:trcase}\includegraphics[width=0.35\textwidth]{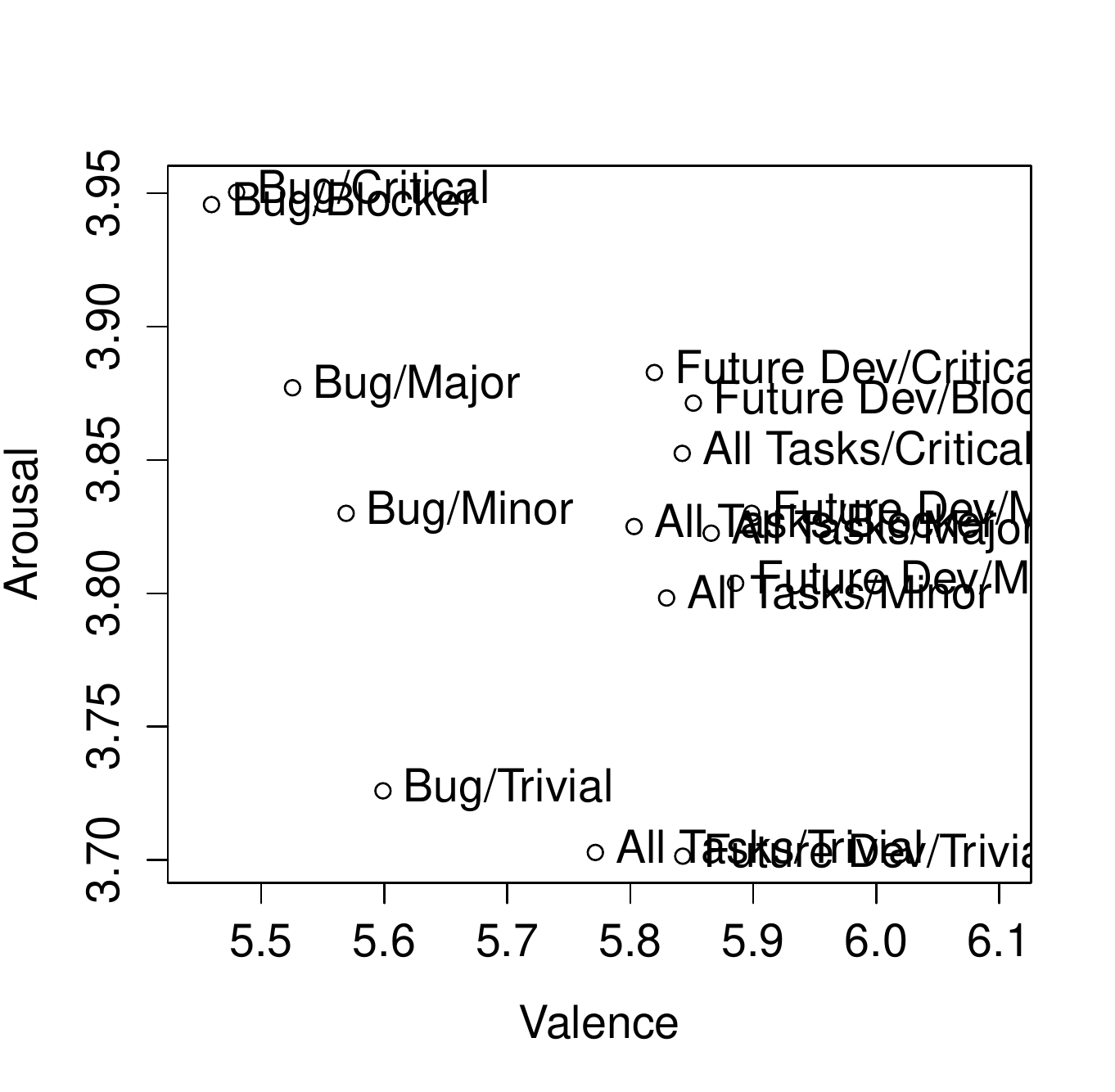}}
\hfil
\subfloat[]{\label{fig:rrcase}\includegraphics[width=0.35\textwidth]{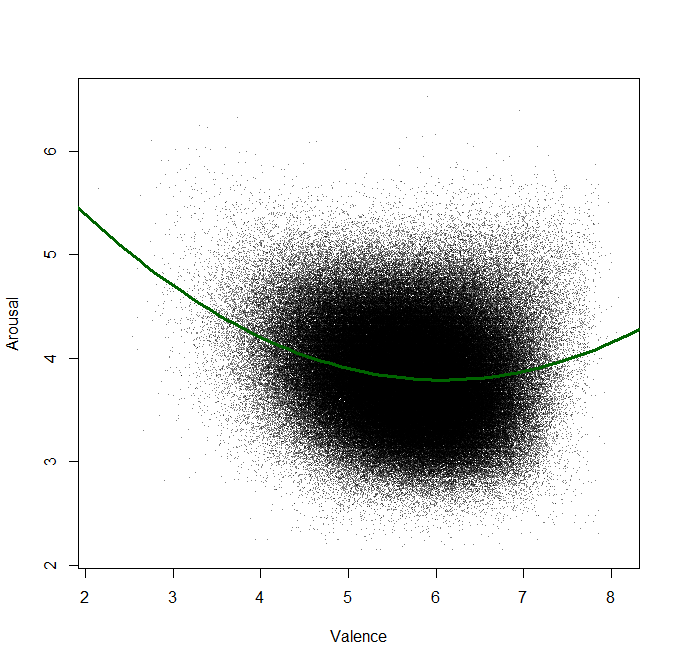}}
\caption{Valence and Arousal (a) averages by Type and Priority and (b) all issues.}
\label{fig:VA_Data}
% \vspace{-15pt}
\end{figure*}

\autoref{fig:VA_Data}(a) summarizes the statistically significant results, i.e., the relationship between issue priority and Arousal and issue type and Valence. For the figure, we used the previously computed scores for Valence, and Arousal, but to simplify things we present only a single score for each VAD dimension instead of having a separate score per issue element. This score is created by averaging the scores given to issue Title, Description and All comments (All comments includes the first and last comment). Additionally, \autoref{fig:VA_Data}(b) shows all issues plotted in the Valence-Arousal space to give a complete view of our data. The line in \autoref{fig:VA_Data}(b) comes from a quadratic model (R$^2$=0.066) that outperforms the linear model when modelling the relationship between Arousal and Valence. There is a theory that the relationship between Valence and Arousal has a U-shaped curve. Words having either very high or very low Valence produce more Arousal than words that are neutral, i.e. having medium Valence.

%%% Local Variables:
%%% mode: latex
%%% TeX-master: "paper"
%%% End:

\subsection*{RQ2: How does VAD change when issues are resolved?}
\textbf{Motivation.} RQ1.3 showed a different relation between Dominance and the last issue comment of a report in comparison to the first issue comment of a report. A potential explanation could be that emotions towards an issue can change over time, i.e., while at the end of an issue report everything is clear and under control, initially Dominance could have been much lower, in which case developers were unsure and hence might have needed more time to resolve an issue.

\textbf{Approach.}  We analyze issues that have been resolved and have at least four comments. This allows us to analyze the VAD scores of the first and last comments. Additionally, we also performed the same analysis for each role involved in an issue report, i.e., Assignee, Report and Other, to understand whether VAD changes depend on the role. For each role, we only required two comments to exist for a given issue.

After filtering our data set for issues with sufficient comments, % we are operating with a smaller subset of the data. O
only 124,537 issues out of 701,002 remain (closed and having 4 or more comments). For the Assignee role, we have 40,322 closed issues where an Assignee has given 2 or more comments. For the Reporter and Other roles the numbers are 25,579 and 51,436, respectively. Thus, our results are only generalizable to issues with sufficient discussion.% Yet, we think the emotional patterns would be similar, albeit smaller due to smaller interest, to other issues as well.  

\textbf{Findings} {\bf From the moment an issue is reported to the time when the issue is closed, Valence and Dominance tend to increase, while Arousal has a small decrease}. \autoref{tab:1stLast_All} shows how for all roles, that Valence at the end of the resolution process of an issue is higher than at the start. The same holds for Dominance, i.e., the issue parrticipants not only feel positive, they also feel in control of the situation. This intuitively makes sense. For Arousal, only a very trivial drop an average is found (d=-0.026). Next, comparison between roles provides interesting additional insights of the changes in emotions.

% Does not hold with new data. Forgot to look at it earlier. 
%More detailed investigation \bram{no figure for this?} learnt that only for trivial issues arousal is considerably higher after an issue has been closed than at the start. Issue reports with high arousal at the start seem to correspond to high priority issues such as ``Blocker Bugs'', suggesting that developers are very active (motivated) when starting to resolve such issues. % Arousal for lower Priority issues increases , or alternatively the Arousal for all issue types and priorities becomes more similar, see .  There are no changes in Arousal of high Priority issues. Interpreting this is straight forward, as an issue gets fixed or is close of being fixed our pleasure increases. 
%The finding for low priority might be due to developers feeling an increase in Arousal because they did not expect the issue to be fixed in the first place. An alternative interpretation is that Arousal for issues getting fixed is similar regardless of issue Priority or Type, but that, for lower Priority issues, one simply started with low Arousal as the issue was not that important to begin with, but once the issue is fixed one still feels a similar emotion of accomplishment. These are hypotheses, and user studies will be necessary to evaluate them in a controlled setting.

{\bf The Arousal of Assignees drops as issues are resolved}. For the Assignees' emotions in \autoref{tab:1stLast_All}, it seems like Arousal drops as time moves closer to issue resolution. This decrease in Arousal seems natural, since, as the issue is being resolved, less and less work is left until the Assignee does no longer need to do anything and hence there is no need to feel active anymore. Valence on the other hand gets a small increase. Surprisingly, the Valence (pleasure) experienced by the Assignee is the lowest one across all four roles (first row). Since the Assignee is the one accomplishing the resolution of an issue, one would expect him or her to be more positive than the other roles. On the other hand, Assignee is the one doing all the work while others simply wait for the Assignee's contribution, thus, the others might experience more Valence as they are the ones that receive the fix from Assignee, i.e., like receiving a gift.

%For Valence, we see less increase in comparison to the general.  We were, however, surprised that less pleasure was measured from Assignee than for other individuals.

{\bf The Valence and Dominance of Reporters increase as issues are resolved, yet Arousal remains stable.} Except for Arousal, the observations for Reporters' emotions in \autoref{tab:1stLast_All} are similar to those for all comments. In comparison to the Assignees and Others,  the Reporters experience the highest increase in Valence and Dominance when their issues are resolved and experience no drop in Arousal.

%In more detailed investigation, we found that low priority issues tend to increase Arousal as issues are resolved, while the high priority ones tend to decrease Arousal. The latter seems intuitive, since towards the end of the resolution process, the original reporter just needs to wait for and test the Assignee's patches. In any case, Arousal has the largest variation and spread in values.% This spread somewhat decreases when time passes, however, in general we can say that Reporters show the largest variation in Arousal.

{\bf For Other commenters' emotions, Valence and Dominance increase as issues are resolved, while Arousal decreases.} \autoref{tab:1stLast_All} indeed shows  a mixture of changes that are partially similar to Assignees' and partial to Reporters' changes. 
We see a drop in Arousal as time passes, similar to what was witnessed for Assignees. Perhaps the Other commenters also feel they are contributing to solving the issue in the comments and as resolution gets closer their Arousal drops as higher activation is no longer needed. On the other hand, Others also get higher increase in Valence and Dominance when issues get resolved.

\subsection*{RQ3: Can VAD explain the time used for fixing a defect?}

\begin{table}[ht!]
\centering
\caption{Significance of the metrics selected in the Logistic regression models containing only control metrics, adding affective metrics and adding VAD metrics. Significant VAD metrics are shown in bold.}
\begin{tabular}{L{2.0cm}lllll}
\midrule
\multicolumn{5}{c}{\textbf{Control Metrics} }\\
\midrule
Estimate                   & Std. Error 	& t value   	& Pr(\textgreater|t|) 	&         \\
\# comments              	&   1.011e-01  	& 7.531e-03  	& \textless 2e-16 		& ***     \\
\# assignee prev. comm.    &   -1.075e-04  & 6.310e-06 	&   \textless 2e-16 	& ***     \\
\# reporter prev. comm.    &   -2.693e-05  & 6.774e-06  	&  7.02e-05 			& ***     \\
\# developers      			&   2.388e-01  	& 1.159e-02  	&  \textless 2e-16 		& ***     \\
\# watchers                	&   2.142e-02  	& 4.575e-03   	&  2.84e-06 			& ***     \\
\# changes         			&   7.737e-02  	& 2.292e-03  	&  \textless 2e-16 		& ***     \\
Critical          	&   3.091e-01  	& 6.964e-02   	&  9.04e-06 			& ***     \\
Major             	&   5.537e-01  	& 4.958e-02  	&   \textless 2e-16 	& ***     \\
Minor             	&   6.988e-01  	& 5.260e-02  	&   \textless 2e-16 	& ***     \\
Trivial           	&   4.642e-01  	& 6.734e-02  	&   5.44e-12 			& ***     \\
\midrule
\multicolumn{5}{c}{\textbf{Affective Metrics} }\\
\midrule
%Estimate                   & Std. Error 	& t value   	& Pr(\textgreater|t|) 	&         \\
AVG sentiment        		&  -4.089e-01  	& 1.052e-01  	&  0.245 			    &         \\
AVG politeness       		&   3.853e-01  	& 3.966e-02  	&    \textless 2e-16 	& ***     \\
AVG love                  	&   -1.118e+00 	& 5.803e-02 	&   \textless 2e-16 	& ***     \\
AVG JOY                   	&   -9.218e-01  & 8.107e-02 	&   \textless 2e-16 	& ***     \\
AVG sadness               	&   2.914e-01  	& 4.537e-02  	&  1.34e-10 			& ***     \\
title politeness          	&   9.874e-02  	& 3.817e-02   	&  0.009688 			& **      \\
first comm. sentiment   	&   1.997e-01  	& 5.803e-02   	&  0.000577 			& ***     \\
last comm. sentiment    	&   2.699e-01  	& 6.651e-02   	&  4.94e-05 			& ***     \\
last comm. politeness   	&   -1.392e-01  & 1.896e-02  	&  2.14e-13 			& ***     \\
\midrule
\multicolumn{5}{c}{\textbf{VAD Metrics} }\\
\midrule
%Estimate                   & Std. Error 	& t value   	& Pr(\textgreater|t|) 	&         \\
\textbf{title A}     &   -1.578e-01 	&   4.723e-02   &   0.014634  			& *	  \\
\textbf{title V}     &    3.968e-01 	&   5.270e-02   &   9.57e-06  			& ***	  \\
desc. A        		&   1.833e-01 	&   1.039e-01   &   0.420120  			& 	  	  \\  
\textbf{desc. V}     &   4.241e-01 	&   1.202e-01   &   0.000139  			& ***	  \\
\textbf{all comm. A} &    1.139e+00 	&   1.920e-01   &   3.01e-09  			& ***	  \\
\textbf{all comm. V} &   -1.734e+00 	&   1.738e-01   &   < 2e-16  			& ***	  \\
all comm. D       	&    5.845e-01 	&   2.269e-01   &   0.102523  			&   	  \\
\textbf{first comm. A}&  1.403e-01  	&   7.518e-02   &   1.69e-07  			& *** 	  \\
first comm. V			&  -1.518e-01 	&   7.844e-02   &   0.32	  			&  	  \\
\textbf{last comm. A}&   -4.764e-01 	&   8.070e-02   &   3.57e-09  			& ***	  \\
last comm. A         &    1.519e-01 	&   7.295e-02   &   0.45    			& *  	  \\
\end{tabular}
\label{tab:logistic_reg_significance}
\end{table}

%\todo{Still old RQ3 need to be update with new results}
\textbf{Motivation.} RQ 1.3 found certain links between VAD (i.e., Dominance) and issue resolution time, while RQ2 showed changes in VAD
between the start and end of an issue report. These findings hint that VAD measures could provide substantial explanatory power of why a given 
defect or feature is taking longer to fix or develop. Since existing research has already explored the links between resolution time and discrete emotions, 
sentiment and politeness~\cite{Murgia2014,Ortu2015a}, here we compare the role of VAD measures to that of these other emotion measures 
in explaining resolution time. The underlying assumptions are that, since (1) discrete emotions like anger, joy and sadness~\cite{Murgia2014} can be mapped to the 
VAD measures and (2) the VAD measures are independent from cultural or linguistic interpretation of words~\cite{guerini2015deep,Russell1991}, the 
VAD measures are more fundamental and should explain issue resolution time better.

\textbf{Approach.} We built a logistic regression model to investigate 
which variables are associated the most with issue resolution time considering
the 14 projects used by Ortu et al.~\cite{Ortu2015a} with 60K issues and 500K comments. The model is built hierarchically, as follows.
First, we build a model using the control metrics used by Ortu et al.~\cite{Ortu2015a}, then a second model is built using both the control and affective
metrics of Ortu et al.~\cite{Ortu2015a}. 
Finally, a model with all control, affective and VAD metrics is built. We build classification models (logistic regression), since we discretized the output variable (issue resolution time) into a binary variable: ``Short'' (resolution time lower than median resolution time) and ``Long'' (resolution time larger than or equal to median resolution time).

To evaluate precision and recall, we use 10-fold cross-validation, which divides the data set into 10 smaller sets, each of which, in turn, is used as test set. We then measure what percentage of issues classified as ``Long'' really are ``Long'' (precision), as well as what percentage of all ``Long'' issues really were classified as such (recall). To compare performance to a random classification model, we also calculated the AUC value. The higher the value is compared to 0.5, the better the model performed compared to a random model. We also compare to a ZeroR model, which is a simple baseline model that always outputs the majority class (``Long'' in our case). The better a model is compared to this baseline, the more useful it is in practice. Note that all these models are explanatory in nature, no prediction is performed.

In order to determine what metrics have the highest impact in (``dominate'') the model, we build one model on the whole data set instead of using 10-fold cross-validation. First, we % do this only with the control metrics, then use
% Given these models, we first 
analyze whether
each set of metrics' model statistically significantly increases upon the previous one, and which metrics are significant. Next, by applying ANOVA analysis on the final model, we remove insignificant metrics to build the final model.
The final model is then used to % predict the issue fixing time and to 
evaluate the impact of VAD metrics on the final model by measuring an impact size~\cite{Ortu2015a}.

This impact size is calculated by first putting all variables in the final logistic model to their median value. The resulting probability is called the $base$ probability. 
Then, one variable at a time, one standard deviation is added to the variable's median (while all other variables are left at their median value) and the resulting probability $dev$ is used to 
calculate the impact size $\frac{dev-base}{base}$. This basically represents the percentage of increase in probability compared to the $base$ probability 
for a typical % change of a metricobtained by putting all variables to their median value. In other words, it expresses the impact of a typical 
increase of a variable in the model. 
Since each variable can use a different unit and have a different variance, one cannot just compare the model's coefficients for these variables to each other.

\textbf{Findings.}
{\bf Adding VAD metrics statistically significantly improves the explanatory models for resolution time.}
We compared the logistic regression models containing only affective metrics to those containing both affective and VAD metrics using ANOVA analysis 
(with a Chi-squared test), obtaining a p-value of \textbf{2.2e-16 ***}. This confirms that these two models are statistically significantly different, 
with the VAD model improving the fit of the models (both models also improved on the initial control model). We investigated the correlation between 
affective and VAD metrics and we only found a significant correlation greater than 0.7 between \textit{Dominance and Valence} of \textit{issue last comment, first comment, title and description}, thus we removed those Dominance metrics from the model.

\autoref{tab:classifier_performance} shows that this improvement also translates itself into better precision, recall, F-score (harmonic mean of precision and recall) and AUC, although the 
increases are not dramatic. For example, the AUC value increases from 0.747 to 0.782, while the F-score increases from 0.717 to 0.75, 
indicating that both precision and recall (slightly) change at the same time.

{\bf Most of the VAD measures for issue title, description and all comments are significant}. \autoref{tab:logistic_reg_significance} shows for each of the control, 
affective and VAD metrics, how significant it is in the logistic regression model. Variables with at least two stars are significant with $\alpha < 0.01$. 
We highlighted those VAD metrics in bold and removed the remaining VAD metrics from the final model.%the remaining VAD metrics.

{\bf Across VAD metrics, Valence of all comments and the Valence  of the issue title have the largest impact on issue resolution time}. Table \ref{tab:affective_metrics_impact} shows the impact of VAD metrics on the logistic regression model, ordered from most extreme to closest to zero (VAD metrics in bold). 
The top metric is Valence of all issues, which has a very negative impact size. In other words, higher Valence across comments reduces the time to resolution. 
This intuitively makes sense, since the more fun an issue seems to be, the faster one ought to progress.

The next two metrics (Valence of issue title and Arousal across all comments) have a positive impact, which means that higher Valence and/or 
being more in control across the issue comments prolong the resolution time of an issue. These impacts are less intuitive. In fact the Valence and 
Arousal impact of all comments has the opposite sign of the corresponding metrics on just the title. It is not clear to us why this happens. 
%The issue report VAD metrics tend to follow those of all comments, while the Arousal on last comment is similar to issue titles' Arousal.

%\todo{missing: impact size of the other metrics, i.e., the goal of the RQ is to compare the impact of VAD to the other affective metrics (and even control metrics), but currently we only have the impact size for VAD metrics ...}

\begin{table}[t] 
\centering
\caption{Logistic regression model performance.}  
\begin{tabular}{m{2.2cm}m{1cm}m{1.2cm}m{1cm}m{0.5cm}m{0.5cm}}    
\toprule
\textbf{Classifier} & \textbf{Time} & \textbf{Precision}& \textbf{Recall} &
\textbf{F1} & \textbf{AUC}\\
\midrule

\multirow{3}{2.5cm}{ZeroR} 	& 	Short	&  0     	&     0 	&    0    	&
\multirow{3}{1cm}{0.5} \\
							&   Long	&  0.565  	&  1  		& 0.722   	&  \\
							&   Weighted Avg.	& 0.319  	&  0.565 	& 0.408 	&  \\
\midrule
\multirow{3}{2.5cm}{Logistic with control metrics only} 	
&   Short			& 0.644	  	& 0.634		& 0.639 	& \multirow{3}{1cm}{0.747} \\
&   Long			& 0.713  	&  0.722	& 0.717 	&  \\
&   Weighted Avg.	& 0.682  	&  0.683 	& 0.683	 	&  \\
\midrule
\multirow{3}{2.5cm}{Logistic with control and affective metrics} 	
& 	Short			& 0.648  	& 0.62 		& 0.634 	& \multirow{3}{1cm}{0.759} \\
&   Long			& 0.708  	&  0.733 	& 0.72	 	&  \\
&   Weighted Avg.	& 0.681  	&  0.683 	& 0.682	 	&  \\
\midrule
\multirow{3}{2.5cm}{Logistic with all metrics} 	
& 	Short			& 0.692  	&  0.621 	& 0.654 	& \multirow{3}{1cm}{0.782} \\
&   Long			& 0.721  	&  0.78 	& 0.75	 	&  \\
&   Weighted Avg.	& 0.708  	&  0.71 	& 0.707	 	&  \\

\bottomrule  \hline
\end{tabular} 
\label{tab:classifier_performance}
\end{table}

\begin{table}[t] 
\centering
\caption{The impact of VAD metrics on issue fixing time.}  
\begin{tabular}{m{4.5cm}L{4cm}} 
\toprule   
\textbf{Feature} 	& \textbf{\% of increase of logistic probability when adding one SD}\\
\midrule
%\multicolumn{2}{c}{\textbf{Control Metrics} }\\
%\midrule

\# comments                   	&  	533.92\%	\\
avg \# sentences   				&   78.33\%	\\
AVG politeness       			&   57.09\%		\\
\textbf{title V}  				& 	40.90 \% \\
\textbf{desc. V}   				& 	28.16 \% \\
\textbf{all comm. A}  			&  	27.31 \% \\
title politeness          		&   16.99\%		\\
title sentiment          		&   16.04\%		\\
\textbf{last comm. A} 			& 	15.24\%  \\
last comm. sentiment    		&  	14.61\%		\\
first comm. sentiment    		&  	10.82\%		\\
first comm. politeness   		&   10.17\%		\\
AVG sadness               		&  	6.46\%		\\
\textbf{first comm. A}			& 	3.00\%  \\
AVG anger               		&  	-15.99\%	\\
\textbf{title A}  				& 	-18.14 \% \\
\# reporter prev. comm.   		&   -32.74\% 	\\
last comm. politeness   		&   -46.31\%	\\
\textbf{all comm. V}  			& 	-51.54 \% \\
AVG love                  		&   -95.13\%	\\
AVG JOY                   		&   -101.99\%	\\
\# assignee prev. comm.    		&   -106.73\%	\\

%\midrule
%\multicolumn{2}{c}{\textbf{Affective Metrics} }\\
%\midrule

%\midrule
%\multicolumn{2}{c}{\textbf{VAD Metrics} }\\
%\midrule

\end{tabular} 
\label{tab:affective_metrics_impact}
\end{table}

\begin{table}[t]
\centering
\caption{The impact of an issue report's characteristics on the VAD scores of its issue comments. Cells with + and - are significant with level <0.001}
\label{tab:predict_VAD}
\begin{tabular}{lccc|ccc|ccc}
                & \multicolumn{3}{c}{Assignee} & \multicolumn{3}{c|}{Reporter} & \multicolumn{3}{c}{Other} \\
                      & V        & A       & D       & V        & A        & D       & V       & A      & D      \\
Priority             & -        & +       & -       & -        & +        &         & -       & +      &        \\
Issue Type                & +        & -       & +       & +        & -        &  +       & +       &        &       \\
Resolution Time & -        & +       &         & -        &  +        &         &         &       &  +      \\
\# votes               & +       &         & +       &          &          &         & -       & +      &        \\
\# comments      & -       & +        & -       & -         & +        &-         & -         & +       &        \\
\# watchers          & -        &        & -       &   +       &          & +       &       &  +      & -      \\
\# ass. prev. iss.            & +       &        &        &          & +        &         &  +       &  -      & +      \\
\# rep. prev. iss.            & -        &         &         &          &  -        &        &         &        &       
\end{tabular}
\end{table}
%%% Local Variables:
%%% mode: latex
%%% TeX-master: "paper"
%%% End:

% \input{RQ3_OLD}

% \input{RQ4}

% \input{RQ5}

 \subsection*{RQ4: What issue characteristics predict VAD scores in the issue comments?}

\textbf{Motivation} As there can be value in predicting (instead of explaining) the emotional states of software developers in terms of risk of productivity loss, we explored which basic variables of an issue can affect VAD scores in the comments. While RQ3 focused on explaining issue resolution time as outcome variable, here we assume that the outside stimuli, such as issue type or even issue fixing time, are the ones that affect VAD scores and not the other way around. Note that even though the resulting models help us understand whether decrease in, say, issue fixing time, coincide with higher Valence, of course the directions of these cause-effect relationships are complex. We can think that pleasure and enjoyment (high Valence) cause shorter defect fixing time as suggested by literature. Yet, it could be that long defect fixing time, due to repeated failed fixing attempts, causes displeasure, i.e., low Valence. Hence, the models only allow to discuss correlation, not causation.%  We think it is likely that the relationship goes both ways and that one cannot say which is the cause and which the effect.

\textbf{Approach} We used linear regression to explore the VAD scores of the different roles involved in issue comments. As we tested nine models (3 VAD scores and 3 roles), we only report the most significant coefficient (<0.001) and whether the relationship increases (+) or decreases (-) a particular VAD score.%s due to lack of space.   

\textbf{Findings.}
{\bf We find that metrics that increase Valence decrease Arousal and vice versa.} \autoref{tab:predict_VAD} shows the relationships that we identified. Here we show two examples for interpreting the table. First, the Assignees' Valence is higher for a low priority issue that is not a bug, has many votes, has a few comments and watchers, high experience of Assignee (\#ass. prev. iss.) and low experience of the issue reporter (\#rep. prev. iss.). The interpretation is that an Assignee is happy when working for an issue that has many votes, i.e., high interest, but does not like watchers or comments as they supposedly reduce his independence or could indicate that there are difficulties. It is not clear why less experienced issue reporters are correlated with increased Valence of Assignees.

As a second example, the assignees' Arousal is higher for high priority issue that are bugs, have a long issue resolution time, and many watchers. This matches very well with the general intuition of high arousal situations: working on something important that needs to be resolved fast, but due to some reason it takes a long time, and at the same time many people are watching. Reporters feel mostly the same way as Assignees, with the difference that their Valence is higher when many people watch the bugs that they have reported, which is opposite to what Assignees feel. 

Dominance often worked in similar fashion as Valence, in line with the literature~\cite{Warriner2013}, but in contrast to Arousal. Thus, it is difficult to characterize the cases of high productivity (high Valence, Arousal and Dominance), as such cases seem rare. For burnout (low Valence, Dominance, and high Arousal) it appears that working on high priority bugs that take a long time to resolve, have many watchers and comments, and which have been reported by an experienced defect reporter might increase the risk of burnout. We found that an assignee's experience increased his/her Valence, whereas for reporters experience decreases Arousal. These two findings suggests that high amount of past experience can reduce burnout risk by either making your experience more pleasure (Valence) or less Arousal. Based on these observations, the information given by \autoref{tab:predict_VAD} could be a starting point for prediction of burnout in a software engineering context. %In particular, combinations of high Arousal and low Valence are risky from the point of view of burnout. Based on our findings, this could be the case for both Assignees and Reporters when an issue has a high number of comments, takes a long time to resolve, is high priority, is bug type, and the person involved has low experience. \bram{repeats previous paragraph}

 %Thus, it is easy to understand why levels of arousal would increase in such scenarios.  

\section{Discussion}
\label{sec:disc}

%In this paper, we have used the dimensional approach to mine emotions in software repositories, in contrast to prior work that used % when the prior works of software engineering repository mining have worked with 
%the discrete approach only. For example, although we have used a general purpose lexicon developed outside the scope of software engineering, it yielded improved model fit in RQ3 on top of the customized emotion and politeness models of % and shown its usefulness. For example,
%Ortu et al. ~\cite{Ortu2015a} % used emotions and politeness scores collected from software engineering context, yet, the VAD measures we used were able to improve the performance of the logistic regression model
%. 

%Furthermore, we have shown links between issue priority and Arousal, as well as issue type and Valence. Both links are motivated from psychology and, thus, their existence verifies that our idea of automatically mining textual software repositories to identify software engineers' emotions seems sound, at least for issue reports. Similarly, our findings show that Valence (pleasure) and Dominance (control) increase as issues are resolved, which again suggest that the approach is sound. Of course, more work is needed to study the VAD of all possible characteristics of issue reports, as well as of other repositories.

We started the paper by proposing that Valence, Arousal, and Dominance information mined from software repositories like issue repositories could offer a way to investigate the productivity and risk of burnout in software developers. Albeit our data is missing a gold-standard, i.e., individual evaluations of their emotions and burnout status, we were able to present several results that support our idea. First, we showed that issue priority increases Arousal, i.e., emotional activation level, measured from issue reports and comments. This is precisely as we expected as more urgent or important tasks should make individuals more aroused. Our results show that this increase in Arousal can also be detected from written issue reports and comments.

Second, we showed that bugs decrease Valence, i.e., the attractiveness of an issue. Again, this is what one would expect, since when something is broken and needs to be fixed the situation is less attractive in comparison to when developing something new. Third, we showed that Valence increases when issues get resolved. In particular, we found that the defect reporters have the largest increase in Valence when their issues are fixed. This resembles a situation where a highly desired achievement, e.g., getting a paper or a grant application accepted, is accomplished, leading to an increase in Valence (attractiveness). Fourth, we found that an assignee experiences a drop in Arousal when an issue gets resolved. Again, this matches common sense, since when an issue or any other type of situation demanding active effort is resolved, there is no longer a need for higher activation.

Fifth, we found that experience might be a factor protecting from burnout risk: less experienced Assignees express lower Valence while less experienced defect reporters express more Arousal. It is easy to understand that higher Arousal appears when one is doing something for the first time (cf. a first public speaking experience). Thus, the results suggest that less experienced individuals are more likely to end up in a situation where high burn-out risk exists, i.e., high Arousal and low Valence. This finding is also supported by literature, see~\cite{maslach2001job}. To summarize, our results show promising initial results, but plenty of future work and improvements are needed, which we will discuss next.

Thus far, we used a simple lexicon-based emotion mining approach. We see some ways of improving our approach. For example, uses of booster words, e.g. "I really like" and negation, e.g. "I do not like", could provide more accurate analysis. However, the lexicon we used provided no scores for such cases. In other words, we do not know how high would be the increase in VAD scores due to booster words. % for example.  

We also assume that if we had used a VAD lexicon tuned for software engineering (SE) the results would be even better as some words carry special meaning in SE. For example, the word ``free'' has high Valence in our general-purpose lexicon. However, in an SE context the word ``free'' is often used in technical discussion of freeing memory and in such case it does not carry emotional meaning connecting to high Valence. This problem of ambiguity and need for SE specific lexicon is also experienced in prior works~\cite{novielli2015challenges,Tourani2014}.

Since we produced results that match with common sense, e.g., priority correlates with higher Arousal and bug issue type with lower Valence, we assume that such an SE-specific lexicon, if available, would produce similar effects, but with higher effect sizes. However, producing such a lexicon is a considerable effort. The lexicon in our study was produced by Warriner et al. using Amazon Mechanical Turk. Perhaps a similar approach could be used as well to produce a lexicon calibrated for a software engineering context.

However, even with an SE-specific lexicon the results would not be perfect. Each project likely would have its own vocabulary and way of using words, hence words might get different meaning even across different projects. Furthermore, % we still would not be perfect as 
words still could have different meanings (and hence VAD scores) in different contexts. As an example, we provide a snapshot of some uses of the previously mentioned word ``free'' from our data. 
\begin{itemize}
\item { ``We build our products warning free (all 5+ million lines), with warnings cranked all the way up.'', indicates high quality, would produce high Valence}
\item { ``i use free software because i dont be leave in software patents'', political statement, amount of Valence depends on whether one agrees or disagrees}
\item { ``I am unable to free up the memory'', technical discussion, should produce medium (not high or low) Valence}
\item { ``Feel free to list obsolete properties here...'', a polite way of expressing need for help, would produce somewhat elevated Valence}
\item { ``Hope to use the free developer edition'', indicates that something has no cost, would produce high Valence}
\item { ``if we stop shipping a war then we are free to do anything we want.'', indicates liberty from past development constraints, would produce high Valence}
\end{itemize}
%Thus, depending on the context a word is used a different VAD score would be required even within in SE domain. 

Additionally, ideas already found in other papers could be used for improvement, especially in the detection of Arousal. For example, in a small study of a single project with two deadlines~\cite{Guzman2013c}, the authors find that, as deadlines came closer, more and longer emails were exchanged with higher emotional intensity. Thus, frequent intervals of exchanging messages and reporting issues could be an additional measure of Arousal. Similarly, Arousal due to time pressure could increase the number of spelling mistakes as it is well-known that under time pressure individuals make speed-accuracy tradeoffs~\cite{kocher2006time}. A more focused discussion also suggests an increase in Arousal. This phenomenon has different names. For example, Mullainathan et al.~\cite{mullainathan2013scarcity} use the word ``tunneling'', while Guerini et al.~\cite{guerini2015deep} talk about ``narrowcasting'' in the context of online new articles. Perhaps, due to focusing, vocabulary with increasing Arousal would contain more names of particular software components and software features.

Finally, we need to point out that the observed effect sizes, for example in RQ1, are small. We think that this only illustrates the general problems in finding emotional responses from text. For example, a highly cited emotion mining paper on Facebook data published in 2014 in the prestigious PNAS journal~\cite{Kramer2014} only reported effect sizes (Cohen's d) ranging from 0.02 to 0.008, whereas our effect sizes are almost 20 times as high (e.g., 0.389).

\section{Threats to validity}

Threats to external validity correspond to the generalizability of experimental results
~\cite{shadish2002experimental}. In this study, we used a dataset containing scores of Valence, Arousal and Dominance (VAD) for 13,915 English words~\cite{Warriner2013}, then we used this lexicon to compute VAD scores for 700,000 issues, and two million comments. 
We considered the dataset proposed by Ortu et al.~\cite{Ortu2015} as a representative sample of the open source world, used in prior works. Yet,
replications on commercial and other open source projects, as well as on other repositories, are needed to confirm our findings.

Threats to internal validity concern confounding factors that can influence the obtained results.
Based on empirical evidence, we supposed a causal relationship between the emotional state of developers and what they write in issue reports~\cite{Warriner2013}. 
Since the main goal of developer communication is the sharing of information, the consequence of removing or disguising emotions may make comments less meaningful and cause misunderstanding. We are confident that the emotions mined were genuine, because the comments used for this work were collected over an extended period from developers not aware of being observed.

Threats to reliability validity correspond to the degree to which the same data would lead to the same results when repeated. 
This research is the first attempt to automatically mine different measures of Valence and Arousal from issue reports, therefore, no previous studies in this field exists to compare our findings.

%Threats to construct validity focus on how accurately the observations describe the phenomena of interest. 
%Mining of emotions from textual issue report comments 
%presents difficulties due to ambiguity and subjectivity. To reduce these threats, we adopted 

%Furthermore [...] measures are approximations and cannot 100\%
%correctly identify the correct affective context, given the challenges of
%natural language and subtle phenomena like sarcasm. 

\section{Conclusion}

This paper contributes to the field of human aspects of software engineering by raising our understanding of how emotions play a role in software development, in particular related to loss in productivity and burn-out. Our study makes five major contributions:

\begin{itemize}
\item To our knowledge, this is the first attempt to use the dimensional approach to emotions in mining software repositories. We do this by utilizing a general purpose VAD lexicon developed in psychology and applying it to a very large repository of 700,000 software issues.
\item We show that increases in issue priority correlate with increases in Arousal, while different issue types impact Valence, so that bug fixing is the least pleasurable.
\item We show that issue resolution correlates with increased Valence, but surprisingly this impact is the smallest for the one resolving the issue (assignee) and the highest for the one requesting the issue to be fixed (issue reporter).
\item We show that VAD score can be used to explain issue resolution time. There is substantial literature showing that increased emotions in terms of VAD correlate with increased productivity, confirming our results.
\item We recognize that the cause-effect relationship between emotions and issue characteristics also go in the opposite direction. An increased issue resolution time was shown to correlate with decreased Valence, but with increased Arousal. Again, this is in line with literature.
\end{itemize}

Most of our findings confirm intuition. For example, if an issue is difficult to solve, then after a set of failed attempts to resolve the issue a drop in Valence will occur. On the other hand, as time passes, the Arousal caused by the need to fix the issue increases as more and more people are affected by the issue and by the fact that the release deadline is likely to get closer and closer. More studies, including on other types of repositories, are necessary to further explore these ideas.

%
% The following two commands are all you need in the
% initial runs of your .tex file to
% produce the bibliography for the citations in your paper.
\bibliographystyle{abbrv}
\bibliography{references}

\end{document}